\documentclass[twocolumn,aps,prr,longbibliography,superscriptaddress]{revtex4-1}
\usepackage{amsmath}
\usepackage{graphicx}
\usepackage{amsmath}
\usepackage{textcomp}
\usepackage[colorlinks=true,bookmarks=false,citecolor=blue,linkcolor=blue,hyperfootnotes=true,urlcolor=blue]{hyperref}
\usepackage[section]{placeins}

\begin{document}
\title{The Role of the Third Dimension in Searching for Majorana Fermions in $\alpha$-RuCl$_3$ via Phonons}
\author{Sai Mu}
\email{sai.mu1986321@gmail.com}\thanks{The first two authors contributed equally}
\affiliation{Materials Science and Technology Division, Oak Ridge National Laboratory, Oak Ridge, Tennessee, 37831, USA}
\affiliation{Materials Department, University of California, Santa Barbara, California 93106, USA}
\author{Kiranmayi D. Dixit }  \email{dixit12@purdue.edu}
\affiliation{Department of Physics and Astronomy, Purdue University, West Lafayette, Indiana 47907, USA}
\affiliation{Quantum Science Center, Oak Ridge, TN 37931 USA.}
\author{ Xiaoping Wang}
\author{Douglas L. Abernathy} 
\author{Huibo Cao} 
\affiliation{Neutron Scattering Division, Oak Ridge National Laboratory, Oak Ridge, TN 37831, USA.}
\author {Stephen E. Nagler} 
\affiliation{Neutron Scattering Division, Oak Ridge National Laboratory, Oak Ridge, TN 37831, USA.}
\affiliation{Quantum Science Center, Oak Ridge, TN 37931 USA.}
\author{Jiaqiang Yan}
\affiliation{Materials Science and Technology Division, Oak Ridge National Laboratory, Oak Ridge, Tennessee, 37831, USA}
\affiliation{Quantum Science Center, Oak Ridge, TN 37931 USA.}
\author {Paula Lampen-Kelley} 
\affiliation{Materials Science and Technology Division, Oak Ridge National Laboratory, Oak Ridge, Tennessee, 37831, USA}
\affiliation{Department of Materials Science and Engineering, University of Tennessee, Knoxville, Tennessee 37996, USA}
\author {David Mandrus}
\affiliation{Materials Science and Technology Division, Oak Ridge National Laboratory, Oak Ridge, Tennessee, 37831, USA}
\affiliation{Department of Materials Science and Engineering, University of Tennessee, Knoxville, Tennessee 37996, USA}
\author{Carlos A. Polanco}
\affiliation{Materials Science and Technology Division, Oak Ridge National Laboratory, Oak Ridge, Tennessee, 37831, USA}
\author{Liangbo Liang}
\affiliation{Center For Nanophase Materials Sciences, Oak Ridge National Laboratory, Oak Ridge, TN 37831, USA}
\author{G\'abor B. Hal\'asz}
\affiliation{Materials Science and Technology Division, Oak Ridge National Laboratory, Oak Ridge, Tennessee, 37831, USA}
\affiliation{Quantum Science Center, Oak Ridge, TN 37931 USA.}
\author{Yongqiang Cheng}
\email{chengy@ornl.gov}     
\affiliation{Neutron Scattering Division, Oak Ridge National Laboratory, Oak Ridge, TN 37831, USA.}
\author{Arnab Banerjee}
\email{arnabb@purdue.edu}
\affiliation{Department of Physics and Astronomy, Purdue University, West Lafayette, Indiana 47907, USA}
\affiliation{Quantum Science Center, Oak Ridge, TN 37931 USA.}
\author{Tom Berlijn}
\email{berlijnt@ornl.gov}
\affiliation{Center For Nanophase Materials Sciences, Oak Ridge National Laboratory, Oak Ridge, TN 37831, USA}

\begin{abstract}
Understanding phonons in $\alpha$-RuCl$_3$ is critical to analyze the controversy around the observation of the half-integer thermal quantum Hall effect. While many studies have focused on the magnetic excitations in $\alpha$-RuCl$_3$, its vibrational excitation spectrum has remained relatively unexplored. We investigate the phonon structure of $\alpha$-RuCl$_3$ via inelastic neutron scattering experiments and density functional theory calculations. Our results show excellent agreement between experiment and first principles calculations. After validating our theoretical model, we extrapolate the low energy phonon properties. We find that the phonons in $\alpha$-RuCl$_3$ that either propagate or vibrate in the out-of-plane direction have significantly reduced velocities, and therefore have the potential to dominate the observability of the elusive half integer plateaus in the thermal Hall conductance. In addition, we use low-energy interlayer phonons to resolve the low temperature stacking structure of our large crystal of $\alpha$-RuCl$_3$, which we find to be consistent with that of the $R\bar{3}$ space group, in agreement with neutron diffraction. 
\end{abstract}

\maketitle 

\section{Introduction}\label{intro} 

In the Kitaev honeycomb model, strong bond-dependent interactions constrain the spins along either the spin-orbital $x$, $y$ or $z$ axis, leading to significant bond frustration ~\cite{kitaev_2006}. This model can be exactly solved to reveal a $Z_2$-type quantum spin liquid state in which there is no long-range magnetic order even down to zero Kelvin. One of the interesting properties of the Kitaev quantum spin liquid is that in zero field, spins can be shown to fractionalize into static and dynamic Majorana fermions, which leads to a spectrum of static $Z_2$ gauge fluxes and itinerant Majorana fermions. In the presence of non-zero magnetic fields, these Majorana and gauge fluxes interact to produce non-abelian bulk anyons and chiral Majorana edge modes, which have potential applications for topological quantum computing~\cite{kitaev_2003,nayak_2008_rmp, aasen_2020_prx, kitaev_2006}. In Ref.~\cite{jackeli_2009, chaloupka_2010} it was proposed that Kitaev interactions can be realized in solid state materials with strong spin-orbit coupling and strong octahedral ligand crystal fields. While the initial search for Kitaev materials was mainly focused on the iridates~\cite{jackeli_2009, chaloupka_2010,liu_2011,choi_2012,feng_2012}, recently,  $\alpha$-RuCl$_3$ has emerged as a very promising candidate material. Signatures of a proximate Kitaev quantum spin-liquid behavior in $\alpha$-RuCl$_3$ have since been reported on the basis of neutron scattering~\cite{banerjee_2016_nmat,do_2017_nphys}, nuclear magnetic resonance~\cite{baek_2017_prl}, terahertz spectroscopy~\cite{wang_2017_prl}, microwave absorption~\cite{wellm_2018_prb}, specific heat measurements ~\cite{widmann_2019_prb}, NMR spectroscopy~\cite {jana_2018_nphys} and Raman spectroscopy ~\cite{mai_2019_prb}. At high fields H $>$ 7.3 T applied in the plane and perpendicular to the Ru bonds, the long-range order disappears leaving behind a quantum paramagnetic state ~\cite{banerjee_2018}. The most straightforward observation of the fractionalization in $\alpha$-RuCl$_3$ was achieved via an observation of the half-integer thermal quantum hall effect~\cite{kasahara_2018_nat} stable to a temperature as high as 5 K. An active research is ongoing to confirm whether $\alpha$-RuCl$_3$ indeed hosts a gapped non-abelian state at high fields~\cite{yokoi_2021_science, czajka_2021_nphys,yamashita_2020_prb,bruin_2021_arxiv}. At the same time, it is being realized that phonons play a crucial role in the observability of the half integer plateaus~\cite {yamashita_2020_prb, ye_2018_prl,aviv_2018_prx}. 

In practical experimental conditions of a thermal Hall effect measurement, the phonons are responsible for transferring thermal energy between the Majorana edge states and the heat baths. As the atoms in $\alpha$-RuCl$_3$ vibrate, their bond lengths and angles change, which dynamically alters the Kitaev interaction, leading to an effective Majorana-phonon coupling. From theoretical studies~\cite{ye_2018_prl,aviv_2018_prx} it was realized that if this Majorana-phonon coupling is not strong enough, the half-integer quantum thermal Hall effect will not be observable even when the Majorana edge states are present. As a first step to quantify the observability of the half integer plateaus, the authors of Ref. ~\cite{ye_2018_prl} derived a formula for thermalization length, defined as the characteristic length scale over which the Majorana edge modes and phonons equilibrate. If the thermalization length is comparative or smaller than the length of the Hall bar, then the Hall conductance will reach the half-integer value. Otherwise, a non-universal smaller value is reached instead. However, to conclude whether $\alpha$-RuCl$_3$ is in the regime for which the half-integer quantum thermal Hall effect is observable, the microscopic properties of the phonons remain to be determined. Of particular importance is the determination of the phonon velocities, i.e. the slopes of the acoustic dispersions towards zero momentum. Phonons with lower velocities are more thermally populated, leading to an enhanced scattering against the Majorana edge modes. Furthermore, for a given energy, the low sound velocity phonons have smaller wave-lengths, meaning larger distortions of the bonds and thus stronger coupling to the Majorana modes. Overall, it was found that the thermalization length is proportional to the fourth power of the phonon velocity~\cite{ye_2018_prl}.      

\begin{figure}[!htb]
\includegraphics[width=0.5\textwidth]{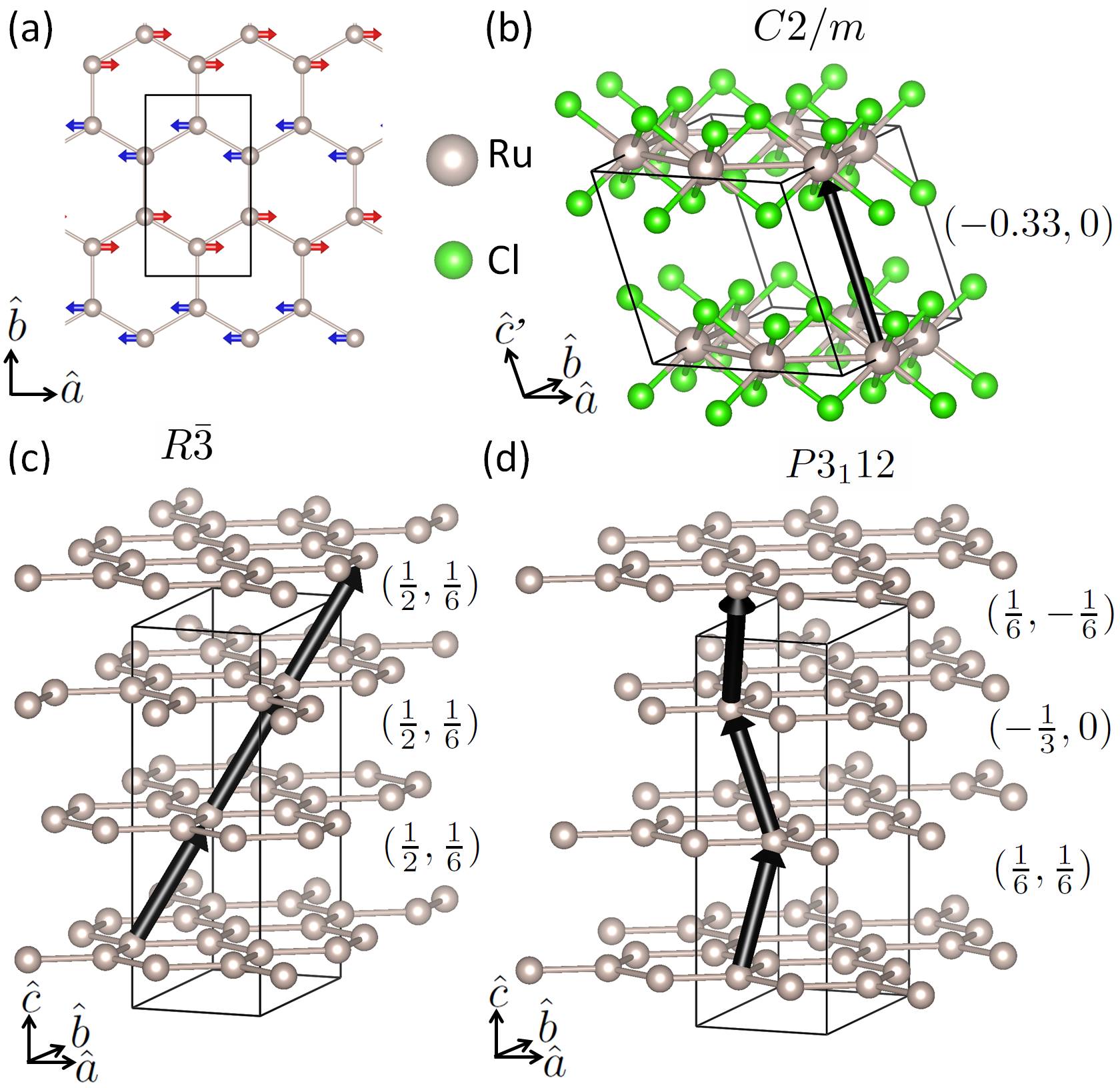}
\caption{\label{fig:fig1} Antiferromagnetic zig-zag unit cells of $\alpha$-RuCl$_3$. (a) In-plane magnetic structure with red and blue arrows denoting anti-parallel spins. Out-of-plane stacking structure for space groups (b) $C2/m$ (No. 12), (c) $R\bar{3}$ (No. 148) and (d) $P3_112$ (No. 151), with black arrows indicating stacking shifts expressed in in-plane lattice vectors. Cl atoms are shown in (b) only.}
\end{figure} 

Phonons can also play a determining role in deducing the interlayer structure of $\alpha$-RuCl$_3$. As shown in Fig. ~\ref{fig:fig1}(b), $\alpha$-RuCl$_3$ is a quasi 2D material which consists of hexagonal layers of Ru$^{3+}$ cations octahedrally surrounded by Cl$^{-}$ anions. Since the RuCl$_3$ layers interact via weak van der Waals forces, they can slide easily over one another. This resulted in the observation of a wide range of space groups that include the trigonal space group $P3_112$~\cite{stroganov_1957, banerjee_2016_nmat,ziatdinov_2016}, the rhombohedral space group $R\bar{3}$~\cite{park_2016,glamazda_2017_prb} and the monoclinic space group $C2/m$ with different proposed stackings~\cite{cao_2016,johnson_2015}. Most larger crystals show a first order structural phase transition around 150 K ~\cite{kubota_2015_prb,cao_2016, ziatdinov_2016,glamazda_2017_prb}, although some smaller crystals do not ~\cite{cao_2016,johnson_2015}. The interlayer structure strongly influences the magnetic properties of $\alpha$-RuCl$_3$. For example, when stacking faults are deliberately introduced, the N\'eel temperature ($T_N$) of $\alpha$-RuCl$_3$ can be increased from 7 to 14 K~\cite{cao_2016}. Importantly, the observation of the half-quantized thermal plateau seems to depend on the interlayer structure, as it is observable in certain samples with $T_N = 7$ K~\cite{yamashita_2020_prb}, but not in others with $T_N = 14$ K~\cite{yokoi_2021_science}. Recently, Raman spectroscopy on low-energy inter-layer phonon modes has emerged as a powerful technique to characterize the interlayer ordering~\cite{liang_2017_acsnano} in quasi 2D materials. The Raman intensities of these so-called shear and breathing modes, in which the layers almost move as rigid objects, are extremely sensitive to the stacking structure, thereby serving as a unique finger print. An interesting question is whether the same concept can be used in neutron scattering experiments to study the interlayer structure in van der Waals materials such as $\alpha$-RuCl$_3$. This is important for the crystals needed for inelastic neutron scattering (INS), where their larger size precludes using X-ray and standard neutron diffraction techniques and confounds the consistency of their structure. 
 
In this article, we perform a comparative study of the phonon structure of $\alpha$-RuCl$_3$ via inelastic neutron scattering (INS) experiments and density functional theory (DFT) calculations. Based on time-of-flight neutron diffraction we conclude that at low temperatures the single crystals used in our experiment conform closely to the $R\bar{3}$ space group. Hubbard U corrections have been included in our phonon simulations as well as the spin-orbit coupling (SOC), the antiferromagnetic (AFM) zig-zag structure and van der Waals (vdW) corrections. Theoretical calculations of the dynamical structure factor are found to be in excellent agreement with our INS experiments. Our experimentally validated first principles force constant model for $\alpha$-RuCl$_3$ is then used to study the low energy phonons relevant for the temperatures at which half-integer quantum thermal hall effect has been reported. Specifically, we analyze the sound velocities and phonon eigenvectors around 0.5 meV. We find that the phonons with their momenta or polarization in the direction perpendicular to the RuCl$_3$ layers have significantly lower velocities compared to the other branches, and therefore that these modes likely will control the observability of the half integer quantum thermal Hall effect in $\alpha$-RuCl$_3$. This can also explain the sensitivity of the half integer quantum thermal Hall effect to the interlayer structure, since the coupling of Majorana fermions to these phonons is naturally sensitive to the interlayer structure of $\alpha$-RuCl$_3$. Furthermore, we also use phonon calculations and measurements to probe the interlayer structure of the same single crystal of $\alpha$-RuCl$_3$. We find that our simulations of out-of-plane phonons based on the $R\bar{3}$ structure are qualitatively in better agreement with our inelastic neutron scattering data then simulations based the $C2/m$ and $P3_112$ structure. We also find that the phonon simulations based on the effective Hubbard repulsion $U_\mathrm{eff}=3.0$ eV better describe the inelastic neutron data than the ones based on $U_\mathrm{eff}=1.0$ eV. 

\section{Methods}\label{methods}

\textit{Neutron diffraction} Single crystals were grown by sublimating pure $\alpha$-RuCl$_3$ powder and were confirmed to have a N\'eel transition at $T_N = 7$ K ~\cite{may_2020_jap}. Single-crystal diffraction was performed using the neutron time-of-flight diffractometer TOPAZ (BL-12) at the Spallation Neutron Source (SNS) at Oak Ridge National Laboratory (ORNL),  which employs a wavelength-resolved technique with a broad 3.1 \AA${}$  neutron bandwidth and an expansive 140 degree in-plane and 32 degree out-of-plane detector coverage providing an expansive reciprocal-space diffraction pattern. A single crystal of $\alpha$-RuCl$_3$ (3.9 mm x 2.6 mm x 0.15 mm) was selected and mounted on a 2 mm diameter Kapton tube on a magnetic base using a minimal amount of epoxy to reduce thermal shear. The sample was mounted on to the TOPAZ goniometer and cooled to 90 K using a laminar flow of liquid nitrogen with an Oxford Cryostream 700 plus cooling system. Care was taken to make sure that the entire crystal was illuminated by the neutron beam which is 4.0 mm in diameter at all times. Data was obtained at both 90 K, 250 K and both during cooling and warming. At 90 K the data was obtained for 35 orientations (40 minutes each orientation) to cover the entire reciprocal space. The data was analyzed using the Mantid software ~\cite{mantid} and an initial model for $R\bar{3}$, $C2/m$ and $P3_112$ were refined against the results of data reduction using SHELX.  Integration ellipsoids~\cite{schultz_2014} used a peak radius of 0.15 r.l.u., inner background radius of 0.16 \AA$^{-1}$ and an outer background radius of 0.19 \AA$^{-1}$ for both Rhombohedral-R and Monoclinic-C cell type and centering analysis.

\textit{Inelastic neutron scattering} Phonon measurements were performed on a 710 mg single-crystal of $\alpha$-RuCl$_3$ at the Wide Angular-Range Chopper Spectrometer (ARCS BL-18) at the SNS at ORNL. The single crystal was mounted on an aluminum shim aligned along the (H0L) axis sealed in an Aluminum soda can with exchange helium gas and inserted into the ARCS standard 5 K bottom-loading cryostat. The data was taken using E$_i$ = 50 meV using the Fermi Chopper 2 (ARCS-100meV-1.5) spinning at 420 Hz and T0 chopper spinning at 60 Hz providing an instrument resolution of $\Delta E = 2.249\pm 0.13$ meV at the elastic line ~\cite{lin_2019_pbcm}. A total of 180 rotation angles at 2-degree spacings, and 0.6 Coulombs proton charge (8 minutes) each, were obtained to symmetrically cover the reciprocal space at 40 K in the rhombohedral phase of $\alpha$-RuCl$_3$ (confirmed via the absence of the (110) Bragg peaks in $R\bar{3}$ notation). Background data was taken with only the aluminum sample holder without a sample, averaged and subtracted at each angle from the sample data. The data analysis was performed using the Mantid routines in the NXSPE format ~\cite{mantid}. The phonon structure derived from the ARCS data is represented in orthogonal momentum-transfer directions with [H,-H, 0], [K, K, 0] in the horizontal plane, and [0, 0, L] out of plane, with reciprocal lattice units 1.21 \AA$^{-1}$, 2.10 \AA$^{-1}$ and 0.037 \AA$^{-1}$, respectively, assuming lattice parameters $a=b=5.975$ \AA{} and $c=17$ \AA{} and indexed in the $R\bar{3}$ space group. The false color images for the momentum vs. energy plots are constructed with a bin size of 0.03 r.l.u. along the [H,-H,0] direction, 0.017 r.l.u. along the [K,K,0] direction, and 0.098 r.l.u. along the [0,0,L] direction. This ensured 0.0363 \AA$^{-1}$ was used for each of the momentum directions unless specifically mentioned otherwise.

\textit{Theory} Density functional theory (DFT) calculations were performed using projector augmented wave (PAW)\cite{Blochl1994} potentials as implemented in the Vienna \textit{Ab-initio} Simulation Package (VASP)\cite{Kresse1993,Kresse1996}. A 500 eV kinetic energy cutoff in the plane-wave expansion has been chosen. The PAW pseudopotentials correspond to the valence-electron configuration 4$d^{7}$5$s^1$4$p^6$ for Ru and 3$s^{2}$3$p^5$ for Cl. The exchange-correlation is treated within the generalized gradient approximation, parameterized by Perdew, Burke, and Ernzerhof (PBE) \cite{Perdew1996}. The Coulomb correlations within the $4d$ shells of Ru were described using the spherically averaged DFT+$U$ method \cite{dudarev1998electron}, where the Hamiltonian only depends on $U_\mathrm{eff}$=$U-J$. We calculated the value of $U_\mathrm{eff}=3.0$ eV for the Ru-$4d$ orbitals from the \textit{ab-initio} linear response method \cite{cococcioni2005linear}. Van der Waals interactions were taken into account via the DFT-D3 method of Grimme \textit{et al.}~\cite{grimme2010consistent}. The in-plane magnetic configuration was stabilized in the experimentally observed antiferromagnetic zig-zag structure (illustrated in Fig.~\ref{fig:fig1}(a)) and spin-orbit interactions were included throughout the calculations. The out-of-plane magnetic configuration for $R\bar{3}$ was taken to be antiferromagnetic between nearest interlayer neighbors in accordance with the experimental observations reported in Ref.~\cite{balz_2021}. In the $C2/m$ simulation the spins nearest along the $\hat{c}'$  direction (see Fig.~\ref{fig:fig1}(b)) were taken to be antiferromagnetic. The out-of-plane magnetic configuration of the $P3_112$ simulation is illustrated in Fig. S7 of the supplement ~\cite{supp}. The initial spin quantization axis was set perpendicular to the Kitaev-Z bond (see Fig.~\ref{fig:fig1}(a)), followed by a full spin relaxation. We note that while the phonon experiments were done above $T_N$, it is important to allow local moments (of the right size) to exists in our phonon simulations. This is because the local moment formation gaps out the electronic density of states near the Fermi level, which in turn weakens the bonds and softens the phonon modes. For practical reasons we do need to treat the magnetic configuration to be ordered, as short-range ordered or disordered moments would require very large supercells. The ZZ AFM configuration observed in experiments is the most natural choice. We also performed non-magnetic phonon calculations. These calculations, however, resulted in imaginary phonon frequencies, similarly as reported in Ref.~\cite{widmann_2019_prb}. For the initial atomic positions of the $R\bar{3}$, $C2/m$ and $P3_112$ structures the lattice parameters in table SII of the supplement ~\cite{supp}, Ref.~\cite{johnson2015monoclinic} and Ref. ~\cite{banerjee_2016_nmat} were used respectively. The structures are illustrated in Fig.~\ref{fig:fig1} (b)(c)(d). Next, the internal atomic positions were relaxed until the maximum force on each atom was less than 5 meV \AA$^{-1}$. The lattice constants and angles were fixed to the experimental values. We note that relaxation of the atomic positions slightly breaks the $R\bar{3}$, $C2/m$ and $P3_112$ symmetries due to the coupling with the antiferromagnetic zig-zag configuration. Harmonic interatomic force constants (IFCs) were calculated using the conventional finite displacement method based on a 48-atom 2$\times$1$\times$2 supercell for $C2/m$ with a 3$\times$3$\times$3 k-point mesh and a 96-atom 2$\times$1$\times$1 supercell for $R\bar{3}$ and $P3_1112$ with a 3$\times$3$\times$2 k-point mesh.  This ensures that the lattice vectors of each supercell exceed 10 \AA. The total energies are converged to 10$^{-8}$ eV per cation. We further enforce the translational, rotational and Born-Huang sum rules by symmetrizing the IFCs following the procedures outlined in Ref. ~\cite{polanco_2020_prm}. With harmonic IFCs, the phonon dispersion relationship at an arbitrary k point is calculated using Phonopy \cite{togo2015first}. From the calculated harmonic IFCs, the phonon dynamical structure factor is simulated using the OCLIMAX code~\cite{cheng2019simulation} with the intensity calculated from a Gaussian-type instrument resolution function determined according to the geometry and other parameters of the beamline. Structural visualizations have been performed using the VESTA software~\cite{momma2011vesta}.

\section{Results} 

\begin{figure}[!htb]
\includegraphics[width=0.5\textwidth]{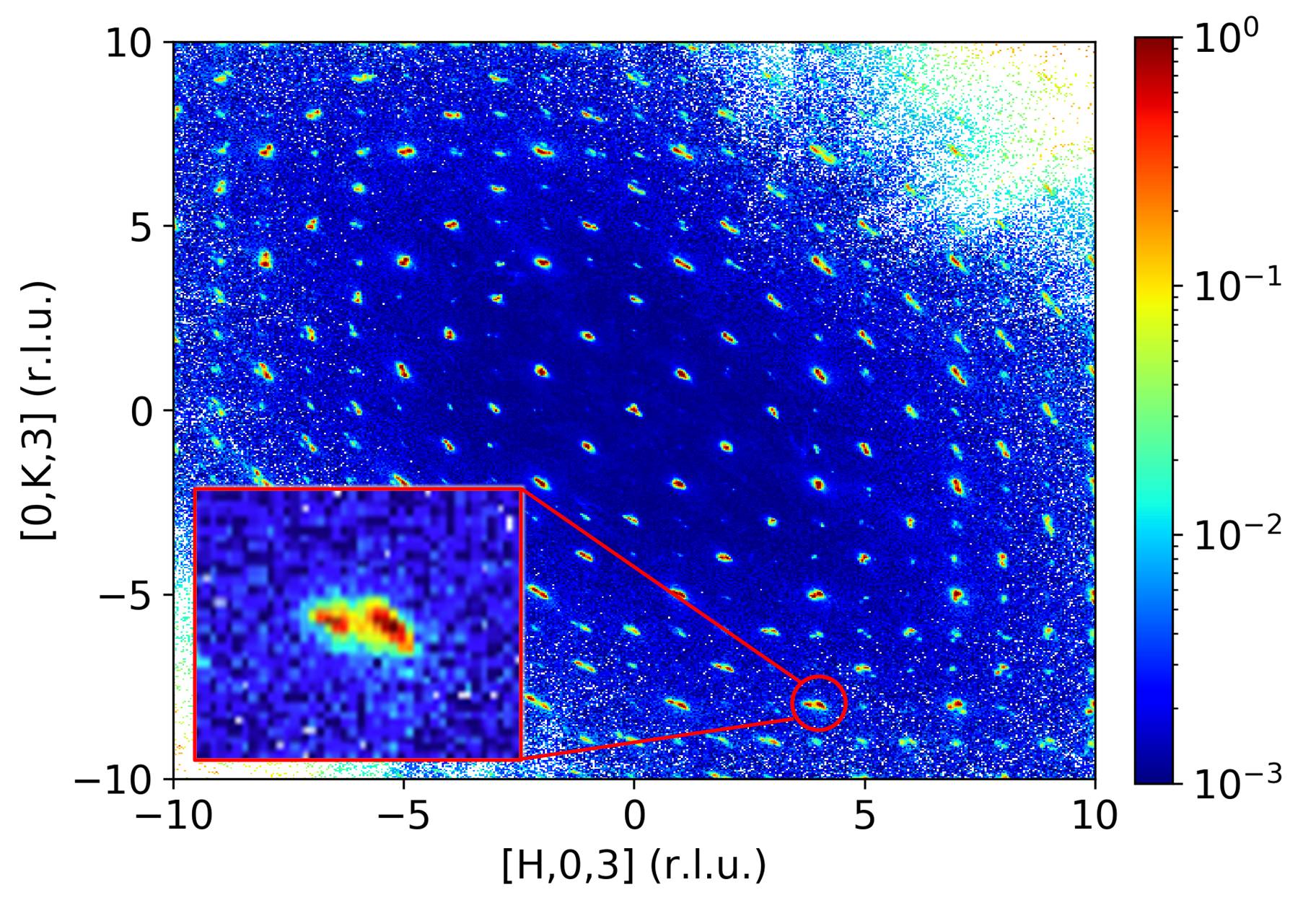}
\caption{\label{fig:fig2} False color plot of in-plane slice of white-beam neutron diffraction  pattern  at 90 K at L $= 3$ (in 0.396 \AA$^{-1}$). Inset (bottom left) shows the radius of integration (see Methods) for the peaks used to extract the $R\bar{3}$ space group structure, which in our case encompasses both the split peaks. }
\end{figure} 

In Fig.~\ref{fig:fig2}(a) and Fig. S1-S3 of the supplemental material ~\cite{supp}, we show the Bragg peaks obtained using the TOPAZ  diffractometer (see Methods). The pattern is broadly consistent with the $R\bar{3}$ space group at 90 K generally matching the results presented in Ref. ~\cite{park_2016}. The crystal chosen is confirmed to contain a low density of stacking faults as verified by the low intensity (by over an order of magnitude) on the forbidden peak positions in Fig. S2.  

We note, however, that the low-temperature space group likely has a slightly lower symmetry compared to $R\bar{3}$. A closer scrutiny reveals an additional peak splitting along the direction H=-K (see inset Fig. \ref{fig:fig2}) below the first order phase transition around 150 K (see Fig. S1). The splitting below the 150 K transition is reversible, conveying that this does not arise from structural damage or twinning of the crystal through the first-order transition (see Fig. S2). This weak symmetry lowering along one direction is likely consequential to the analysis of the structure and the energetics of the sub-$T_N$ zig-zag ground state, and may be related to the observation of the 2-fold symmetric structure in terahertz~\cite{little_2017_prl},  magnetic susceptibility ~\cite{kelley_2018_prb} and magnetotropic coefficient ~\cite{modic_2021} measurements. However, given the inelastic data resolution, we believe this detail would not affect the 90 K phonon data analysis which is the focus in this manuscript. For this purpose, we integrated the structural Bragg peaks using integration ellipsoids which encompassed the split peaks (Fig. \ref{fig:fig2} bottom inset, also see Methods for details on integration radius). Within this caveat, the R-factor that we obtained from the refinement for $R\bar{3}$ is 0.0459, while attempts to index the majority of the peaks in other space-groups yielded an R-factor of 0.1667 for $C2/m$ and 0.2747 for $P3_112$, even when twins were included. The lower value of the R-factor for the $R\bar{3}$ space group points to a significantly better match for our single crystal of $\alpha$-RuCl$_3$ at 90 K. The structural parameters resulting from our refinement at 90 K for $R\bar{3}$ $\alpha$-RuCl$_3$ are given in Table SI of the supplement ~\cite{supp}. The data at 250 K conforms to the $C2/m$ space group which is also seen in smaller crystals. Here we note that the experimental data and the results are also broadly consistent with the results in a previous study ~\cite{cao_2016} once the suitable transformation is used between the $C2/m$ and $R\bar{3}$ space groups. The data analysis in that manuscript was affected by a sign error in the symmetry transformation matrix relating different domains in the $C2/m$ and $R\bar{3}$ structures. Refinements of that data using the correct transformation are consistent with the crystal structure reported here. This also affects the details of the refinement of the ordered magnetic structure below $T_N$ ~\cite{cao_2016}, and will be addressed in a separate manuscript [Cao et al.].

\begin{figure*}[!htb]
\includegraphics[width=1.0\textwidth]{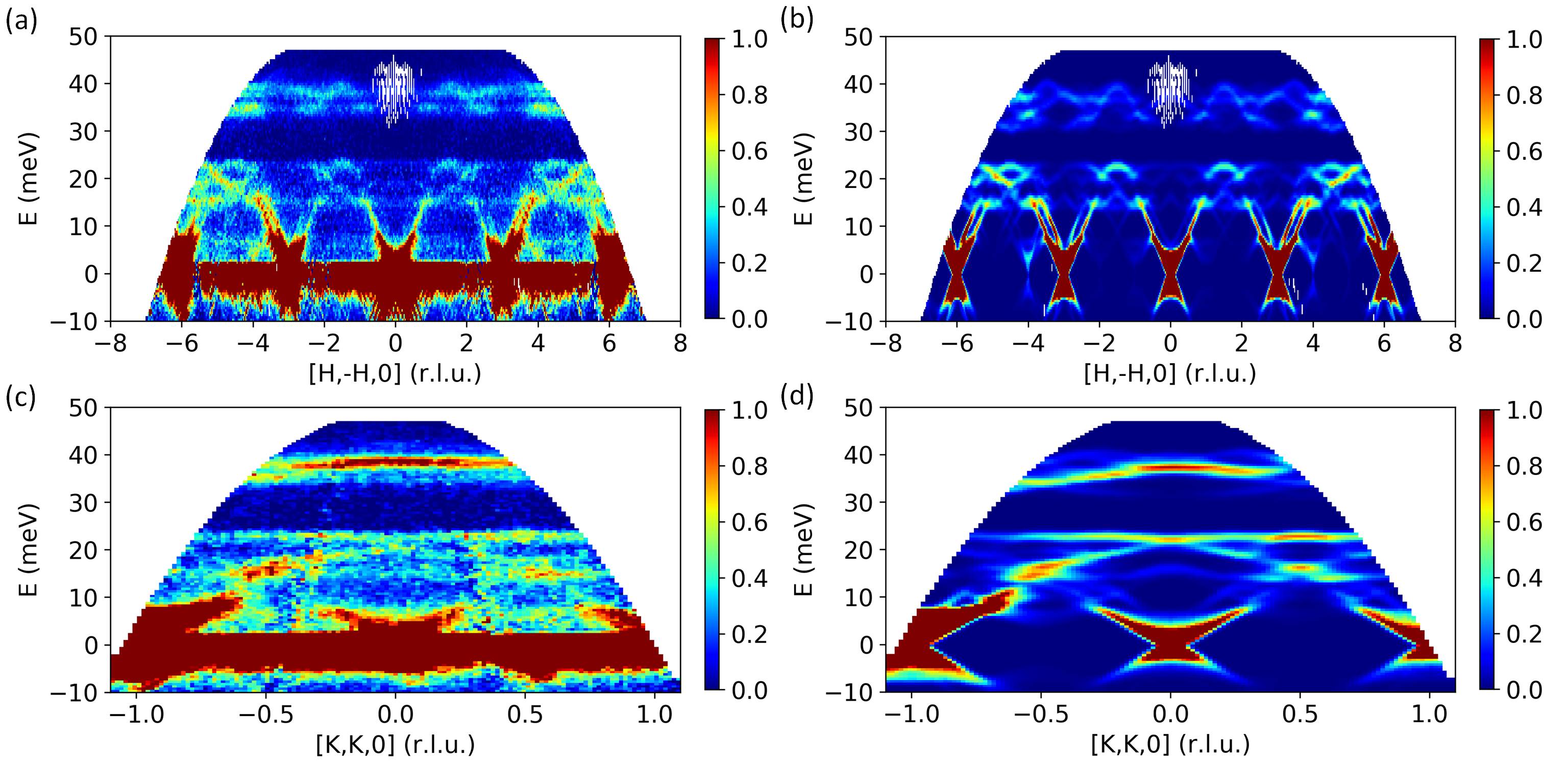}
\caption{\label{fig:fig3} False color plot of the phonon dynamical structure factor of single crystal $\alpha$-RuCl$_3$, measured at 40 K using inelastic neutron scattering with momenta in reciprocal lattice units (r.l.u.) along [H,-H,0] and [K,K,0], respectively. The integration ranges for (a) are L in [11.431, 12.569] r.l.u., and K in [-0.1, 0.1], while for (c) are H in [-0.174, 0.174] r.l.u., and L in [15, 18] r.l.u.  (b),(d): Same as (a) and (c) but from first-principles calculations based on the $R\bar{3}$ space group.}
\end{figure*} 

Having refined the lattice structure of our single crystal sample, we proceed by investigating the phonons of $\alpha$-RuCl$_3$. The phonon band structure of $\alpha$-RuCl$_3$ has thus far been experimentally studied via Raman scattering~\cite{sandilands_2015_prl,sandilands_2016_prb,zhou_2019_jpcs, glamazda_2017_prb,mai_2019_prb}, terahertz spectroscopy~\cite{reschke_2017_prb,reschke_2019_prb}, neutron scattering~\cite{banerjee_2016_nmat,do_2017_nphys}, X-ray spectroscopy~\cite{lebert_2020_jpcm,li_2021_ncomm,hxli_2021_arxiv} and \textit{ab initio} calculations~\cite{sarikurt_2018,widmann_2019_prb}. Here we report a comprehensive study of the phonon structure resolved in the full momentum and energy range by combining Inelastic Neutron Scattering (INS) from the Wide Angular-Range Chopper Spectrometer (ARCS) and density functional theory. In Fig.~\ref{fig:fig3} we present the phonon dynamical structure factor, $S(q,\omega)$, obtained from INS and first-principles simulations in which the lattice parameters from our $R\bar{3}$ refinement were used. The phonon dynamical structure factor calculations are based on simulations in which the moments were initialized to be in the antiferromagnetic zig-zag configuration shown in Fig. ~\ref{fig:fig1} (see methods for more details). Furthermore, the Hubbard Coulomb repulsion among Ru-$4d$ orbitals was taken to be $U_\mathrm{eff}=3.0$ eV. This was the value we obtained from the \textit{ab-initio} linear response method \cite{cococcioni2005linear}. In Fig. S5 of the supplement ~\cite{supp} we also show phonon dynamical structure factor calculations based on $U_\mathrm{eff}=1.0$ eV which we find to agree considerably less with the experimental data. In Fig.~\ref{fig:fig3} we consider the two orthogonal in-plane directions [H,-H,0] and [K,K,0]. Below, in Fig. ~\ref{fig:fig5}, we also consider the third, out-of-plane direction [0,0,L]. As described in the method section, large momenta offsets away from the zeroth Brillouin zone have been used to suppress the magnetic contributions to $S(q,\omega)$. Upon comparing the experimentally observed and theoretically simulated phonon spectra, there are some subtle mismatches. For instance, the gap in the phonon structure, roughly between 25 and 33 meV, is slightly shifted towards lower energy in the theory compared to that of the experiment (c.f. Fig. \ref{fig:fig3}(a)(c)). We also note that the band splitting around [0.5,0.5,0] and 38 meV in the INS, is not captured by the simulations(c.f. Fig. \ref{fig:fig3}(b)(d)). Nonetheless, the DFT simulations reproduce the majority of the complex features in the lower energy section of the INS spectra nearly perfectly within the resolution of the experiment. For example, we attract the attention of the reader to Fig.~\ref{fig:fig3}(a)(b), where we see that the simulation describes remarkably well the intricate details of the set of bands that lies roughly within 14 and 25 meV  with the momentum index H ranging from  4 to 6 r.l.u. Or vice versa by comparing Fig.~\ref{fig:fig3}(c)(d), we see that the band crossing at around 21 meV and ${\rm K}=0$ in the simulation can be recognized in the experimental data, albeit with a small gap. Overall, this represents an excellent agreement between the comprehensive experimental observations of the phonon structure and the unbiased first principles simulations without adjustable parameters. 

\begin{figure}[!htb]
\includegraphics[width=0.48 \textwidth]{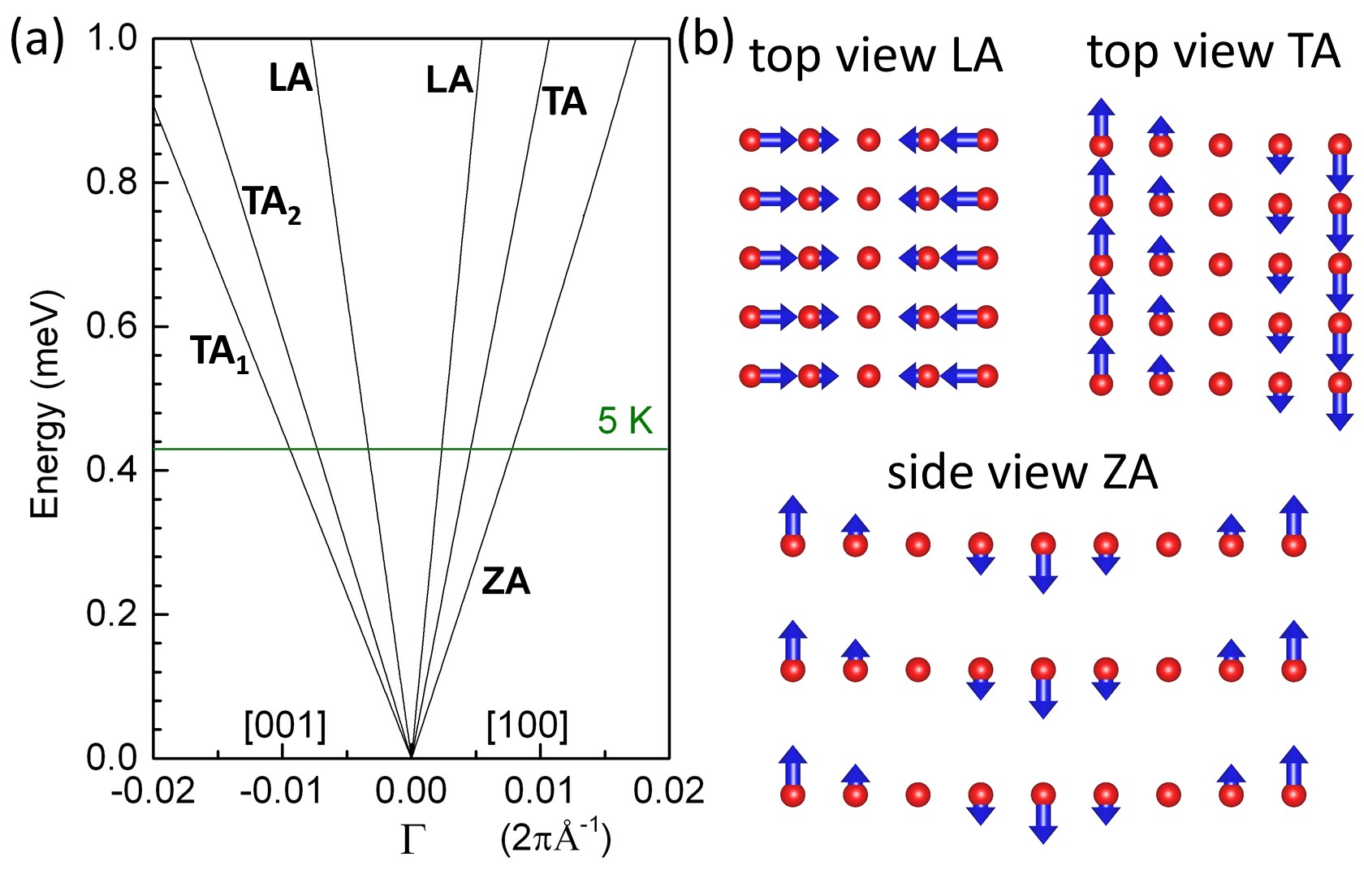}
\caption{\label{fig:fig4}(a) Low-energy phonon bands for antiferromagnetic zig-zag (ZZ) configuration and (b) cartoon of in-plane propagating LA,TA and ZA mode.}
\end{figure}  

\begin{table}[!htb] 
\caption{ Sound velocities of $\alpha$-RuCl$_3$ from first-principles calculations based on the $R\bar{3}$ space group in units of km/s (41.36 meV \AA{}) along crystallographic directions $\hat{a}$,  $\hat{b}$, $\hat{c}$ of the zig-zag $R\bar{3}$ unit cell shown in Fig.~\ref{fig:fig1}.  } \label{tab:tab1}
\begin{tabular*}{.7\linewidth}{@{\extracolsep{\fill}}ccc}\toprule 
 $\vec{q} \parallel \hat{a}$ & $\vec{q} \parallel \hat{b}$& $\vec{q} \parallel \hat{c}$  \\\colrule 
1.32 (ZA) & 1.10 (ZA) & 1.10 (TA$_1$) \\\colrule 
2.26 (TA) & 2.30 (TA) & 1.43 (TA$_2$) \\\colrule 
 4.43 (LA) & 4.39 (LA) & 3.11 (LA) \\\botrule 
\end{tabular*}
\end{table}

Next, we analyze the phonon sound velocities in $\alpha$-RuCl$_3$ from our first principles derived force constants, considering that we have validated these against experiments. The values of the sound velocities are an important factor determining whether a material is in the strong coupling regime in which the debated half-integer quantum thermal Hall effect can be experimentally observed ~\cite{ye_2018_prl,aviv_2018_prx}. In Fig.~\ref{fig:fig4}(a), we show the low energy phonon dispersion along the $\hat{a}$ direction of the AFM zig-zag cell shown in Fig.~\ref{fig:fig1}, which coincides with the [K,K,0] direction in the dynamical structure factor plots shown in Fig. ~\ref{fig:fig3}. The phonon band structure in the full energy range is shown in Fig. S6 in the supplement~\cite{supp}. Here, we focus on the phononic structure below 0.43 meV which corresponds to the temperature scale of 5 K below which the half-integer quantum thermal Hall effect has been reported. We extract the sound velocities by performing a linear fit of the theoretical phonon dispersions within the [0,0.43] meV energy range. The results are summarized in Table ~\ref{tab:tab1}. We note that the average in-plane LA velocity of 4.41 km/s, the average in-plane TA velocity of 2.28 km/s, and the average out-of-plane TA velocity of 1.27 km/s obtained from our first-principles calculations agree well with the corresponding values of 3.49, 2.58 and 1.67 km/s obtained from inelastic X-ray spectroscopy ~\cite{ hxli_2021_arxiv}. Here we multiplied the velocities in Ref. ~\cite{hxli_2021_arxiv} by a factor $2\pi$ to account for the different definition of the crystal momentum. In the model calculations of Ref.~\cite{ye_2018_prl}, for simplicity,  it was assumed there is a single sound velocity for $\alpha$-RuCl$_3$. However, in real materials the sound-velocities of the transverse acoustic (TA) modes are significantly lower than those of the longitudinal acoustic (LA) modes. Furthermore, in quasi-2D materials like $\alpha$-RuCl$_3$ we can differentiate the so-called ZA modes which propagate in the plane, but vibrate in the out-of-plane direction (see Fig.~\ref{fig:fig4}(b)). The ZA modes naturally have a lower energy than the TA phonons due to the out-of-plane restoring forces being weak. Table ~\ref{tab:tab1} shows that the sound velocities in $R\bar{3}$ $\alpha$-RuCl$_3$ strongly vary not only depending on whether the propagation direction is in-plane or out-of-plane, but also on whether the mode index is LA,TA or ZA. Tables SIII and SIV of the supplement ~\cite{supp} shows that the sound velocities of $C2/m$ and $P3_122$ $\alpha$-RuCl$_3$ are quite similar. In addition to the phonon eigenvalues, the phonon eigenvectors also play an important role because these determine the coupling to the Majorana fermions. Although, at higher energies the phonon eigenvectors are quite complex, we find that the phonons within the relevant [0,0.43] meV energy range are nearly perfectly acoustic (see Fig. S8 of the supplement ~\cite{supp}). This elementary structure of the low energy phonon eigenvectors simplifies the analysis of the Majorana-phonon coupling.  

\begin{figure}[!htb]
\includegraphics[width=0.5\textwidth]{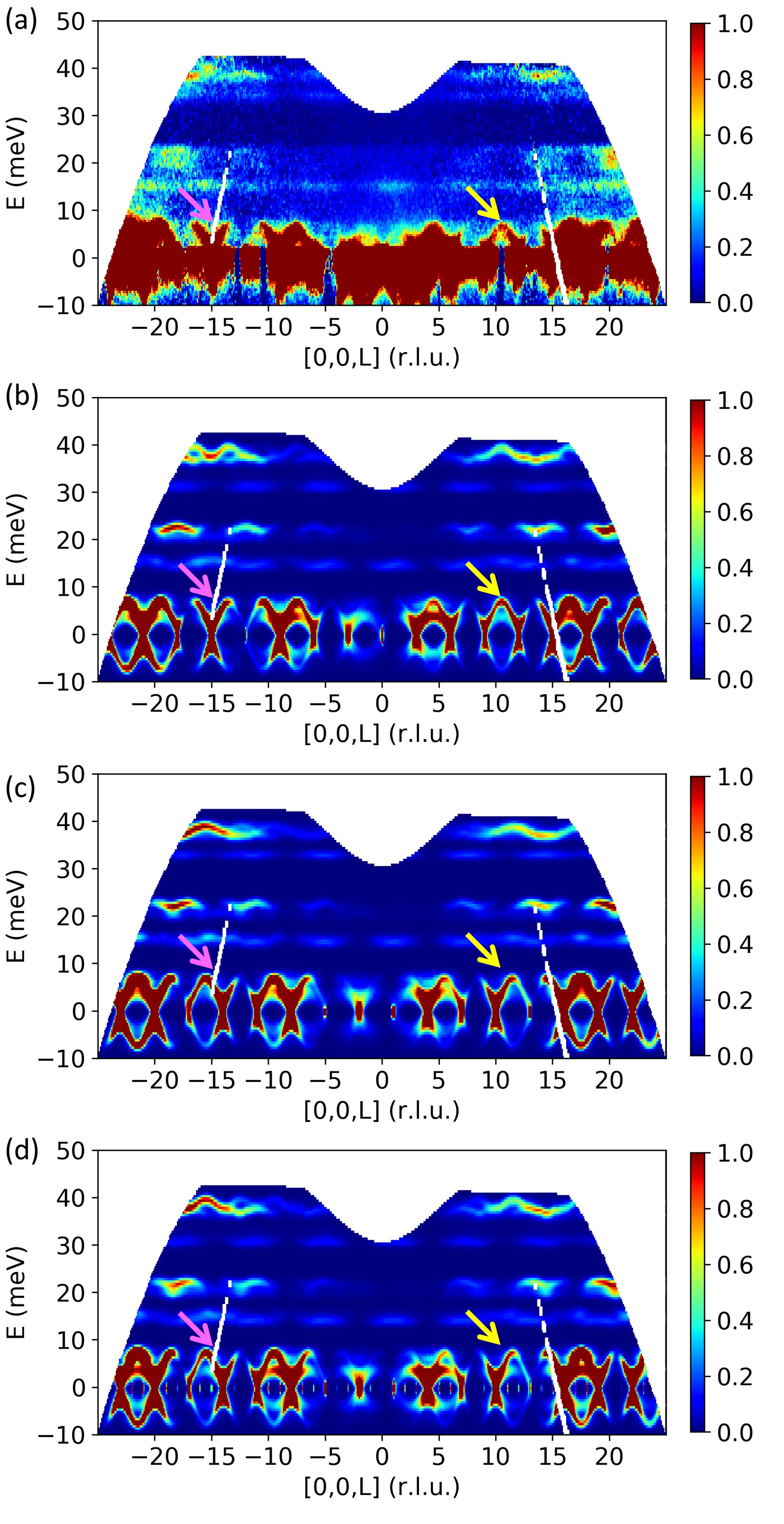}
\caption{(a) False color plot of the phonon dynamical structure factor of single crystal $\alpha$-RuCl$_3$, measured at 40 K from inelastic neutron scattering with momenta in reciprocal lattice units (r.l.u.) along [0,0,L]. The integration ranges are H in [1.326, 1.674] r.l.u., and K in [0.4, 0.6] r.l.u. (b), (c) and (d): same as (a) but from first-principles calculations based on the $R\bar{3}$, $C2/m$ and $P3_112$ space groups respectively.} \label{fig:fig5}  
\end{figure} 

Finally, we investigate the phonons with out-of-plane momenta. In large INS samples, the inevitable existence of domains impedes structural refinement using  X-ray and standard neutron diffraction techniques. This a concern, especially for van der Waals materials like $\alpha$-RuCl$_3$ in which the stacking structure between different crystals can easily vary. In our introduction, we therefore raised the question whether probing low-energy inter-layer phonons via INS can be used to distinguish the stacking pattern in layered materials such as $\alpha$-RuCl$_3$, similarly as it is done via Raman scattering ~\cite{liang_2017_acsnano}. In Fig. ~\ref{fig:fig5} we present the out-of-plane phonon dynamical structure factor obtained from INS. In addition we present the phonon dynamical structure derived from first principles calculations based on the $R\bar{3}$, $C2/m$ and $P3_112$ space groups. Each of these space groups represent a different stacking pattern as described in Fig ~\ref{fig:fig1}. Upon comparing the theoretical and experimental correlation functions, we see that the simulated $S(q,\omega)$ based on the $R\bar{3}$ stacking structure, agrees well with the INS data. For example, if we focus on the pink arrows on the left, we see that both the neutron scattering data and the $R\bar{3}$ simulation consistently show a minimum in the phonon dispersion around 8 meV and $L=-15$, as opposed to the maximum of the phonon bands displayed in the $C2/m$ and $P3_112$ simulations. As an opposite example, we can consider the region at the same energy and the momenta given by $L=10$, indicated by the yellow arrows on the right. Here we see that the phonon dispersions obtained from the neutron experiment and the $R\bar{3}$ simulation are at a maximum. The phonon bands obtained from the $C2/m$ and $P3_112$ simulations on the other hand are somewhere between a maximum and a minimum instead. We note that the simulated out-of-plane phonon bands for the $C2/m$ and $P3_112$ stackings are remarkably similar. We can ascribe this to the fact that the stacking shift of $P3_112$ is essentially the same as that of $C2/m$, albeit with a 120 degree rotation in going from one van der Waals gap to the next (see Fig. ~\ref{fig:fig1}(b)(d)). On the whole, we find that for the momenta perpendicular to the RuCl$_3$ planes, the phonon simulations based on the $R\bar{3}$ space group agree much better with the inelastic neutron data than the ones based on the $C2/m$ and $P3_112$ space groups. Additionally, in Fig. S4 in the supplement ~\cite{supp} we show that the low-energy acoustic phonon bands within the HK momentum plane obtained from the $R\bar{3}$ simulation agree better with experiment than the ones obtained from the $C2/m$ and $P3_112$  simulations. Our conclusions based on the dynamical structure factor are consistent with our neutron diffraction observations, which also indicates that $\alpha$-RuCl$_3$ is better refined by the $R\bar{3}$ space group than the $C2/m$ and $P3_112$ space groups. 

\section{Discussion}

Theoretical studies have shown that the half-integer quantum thermal Hall effect is observable only when the length scale over which Majorana edge modes thermalize with bulk acoustic phonons is equal to or smaller than the length of the Hall bar ~\cite{ye_2018_prl,aviv_2018_prx}. In particular it has been found that in two dimensions the thermalization length depends on the fourth power of the phonon velocity ~\cite{ye_2018_prl}. This power will even raise to five in three dimensions due to the cubic dependence of the phonon density of states on the phonon velocity. In our work, we have shown that in $\alpha$-RuCl$_3$ the phonon velocities, and therefore the thermalization lengths, depend strongly on the polarization of the phonons and the direction of their momenta. For example, in Table ~\ref{tab:tab1} we see that along the $\hat{b}$ direction, the velocity of the ZA mode is roughly a factor 2.1 smaller compared to the TA mode. This implies a reduction of the thermalization length by a factor of $2.1^5\sim 40$. Overall, we find that the out-of-plane TA modes and the in-plane ZA phonons have the lowest velocities, and therefore that these phonons have the potential to dominate the observability of half integer plateaus in the thermal Hall conductance. This leads to an interesting question. If the dominating low-velocity phonons either propagate or vibrate in the out-of-plane direction, could this explain why the half integer quantum thermal Hall effect breaks down in the presence of stacking faults~\cite{yokoi_2021_science}? The coupling of the Majorana modes to the out-of-plane vibrating or propagating phonon modes naturally will be sensitive to the stacking structure. In this respect, our results throw a new light on the difficulty for the half-integer plateau observations to be reproduced, and also offer a path forward, namely control over the interlayer structure. Our findings also indicate that the renormalization of the phonons due to Majorana fermions ~\cite{metavitsiadis_2020_prb,ye_2020_prr, feng_2021_arxiv, metavitsiadis_2021_arxiv} is enhanced for the phonons with their momenta or polarization in the out-of-plane direction, and that it could therefore be sensitive to the stacking structure of the crystals. These conclusions on the role of the third dimension in the search of Majorana fermions via phonons are relevant not only for $\alpha$-RuCl$_3$, but for van der Waals based Kitaev candidate materials in general ~\cite{katoaka_jpsj_2020,lee_prl_2020, ji_2021_cpl}, a class of materials that offer the potential of heterostructure engineered Kitaev quantum spin liquids, and true two-dimensionality via exfoliation. 

To study the interlayer structure of $\alpha$-RuCl$_3$ we have used inelastic neutron scattering from low-energy interlayer phonons in addition to neutron diffraction. This led us to conclude that the low temperature interlayer structure in the large single-crystals used in our neutron experiments is consistent with that of the $R\bar{3}$ space group, and not the $C2/m$ and $P3_112$ space groups.  For $\alpha$-RuCl$_3$ and van der Waals materials in general, the stacking structure can display strong variations. We therefore, consider our ability to verify the interlayer structure in large crystals directly from the inelastic results to be an important and novel result. Although the stacking pattern mainly influences the interlayer magnetic structure, it also modifies the bond lengths and angles within the $\alpha$-RuCl$_3$ planes which could in turn have significant effects on the intra-plane low-energy magnetic properties.  Most first principles derivations of the magnetic exchange couplings have been based on the $C2/m$ and $P3_112$ space groups ~\cite{kim_2016,winter_2016,yadav_2016, hou_2017, eichstaedt_2019_prb}. It will be important to revisit these derivations using the $R\bar{3}$ space group. Furthermore, our study calls for future neutron  experiments to obtain higher resolution  data to discern the subtle symmetry lowering below the $R\bar{3}$ space group at low temperatures. Such studies will be highly relevant for understanding the full scale of anisotropies in the magnetic Hamiltonian of $\alpha$-RuCl$_3$ that leads to both the emergence of the long-range zig-zag antiferromagnetic order and potentially the Kitaev quantum spin liquid phase.

\acknowledgements 

The authors are grateful to Satoshi Okamoto, Pontus Laurell, Mengxing Ye, Simon Th\'ebaud, Peter Czjaka and Lucas Lindsay for fruitful discussions, and to Zach Morgan for his help with the extraction of the diffraction data. The work by TB was conducted at the Center for Nanophase Materials Sciences, which is a DOE Office of Science User Facility. The work by AB, KD, SEN and GBH has been supported by the U.S. Department of Energy, Office of Science, National Quantum Information Science Research Centers, Quantum Science Center. The work by SM and JY was supported by the U.S. Department of Energy, Office of Science, Basic Energy Sciences, Materials Sciences and Engineering Division. KD was also supported by Purdue University, College of Science, Ralf Scharenberg Fellowship. DM acknowledges support from the Gordon and Betty Moore Foundation’s EPiQS Initiative, Grant GBMF9069. A portion of this research used resources at the Spallation Neutron Source, a DOE Office of Science User Facility operated by the Oak Ridge National Laboratory. This research used resources of the National Energy Research Scientific Computing Center (NERSC), a U.S. Department of Energy Office of Science User Facility operated under Contract No. DE-AC02-05CH11231. This research also used resources of the Compute and Data Environment for Science (CADES) at the Oak Ridge National Laboratory, which is supported by the Office of Science of the U.S. Department of Energy under Contract No. DE-AC05-00OR22725.

This manuscript has been authored by UT-Battelle, LLC under Contract No. DE-AC05-00OR22725 with the U.S. Department of Energy. The United States Government retains and the publisher, by accepting the article for publication, acknowledges that the United States Government retains a non-exclusive, paid-up, irrevocable, world-wide license to publish or reproduce the published form of this manuscript, or allow others to do so, for United States Government purposes. The Department of Energy will provide public access to these results of federally sponsored research in accordance with the DOE Public Access Plan (http://energy.gov/downloads/doe-public-access-plan).

\end{document}